\documentclass[amsmath,amssymb,prd]{revtex4}
\usepackage[]{graphicx}

\begin{document}

\title{Quantum computing in a piece of glass}

\author{Warner A. Miller}
\email{wam@fau.edu}
 \affiliation{Department of Physics, Florida Atlantic University,  777 Glades Road, Boca Raton, Florida 33431}

\author{Paul M. Alsing}
\affiliation{Air Force Research Laboratory, Information Directorate, 525 Brooks Road, Rome, NY 13441}

\author{Grigoriy Kreymerman} 
 \affiliation{Department of Physics, Florida Atlantic University,  777 Glades Road, Boca Raton, Florida 33431}

\author{Jonathan R. McDonald}
\affiliation{Air Force Research Laboratory, Information Directorate, 525 Brooks Road, Rome, NY 13441}

\author{Christopher Tison}
 \affiliation{Department of Physics, Florida Atlantic University,  777 Glades Road, Boca Raton, Florida 33431}

\begin{abstract}
Quantum gates and simple quantum algorithms can be designed utilizing the diffraction phenomena of a photon within a multiplexed holographic element. The quantum eigenstates we use are the photon's linear momentum (LM) as measured by the number of waves of tilt across the aperture. Two properties of quantum computing within the circuit model make this approach attractive. First, any conditional measurement can be commuted in time with any unitary quantum gate - the timeless nature of quantum computing. Second, photon entanglement can be encoded as a superposition state of a single photon in a higher-dimensional state space afforded by LM. Our theoretical and numerical results indicate that OptiGrate's photo-thermal refractive (PTR) glass is an enabling technology. We will review our previous design of a quantum projection operator and give credence to this approach on a representative quantum gate grounded on coupled-mode theory and numerical simulations, all with parameters consistent with PTR glass. We discuss the strengths (high efficiencies, robustness to environment) and limitations (scalability, crosstalk) of this technology. While not scalable, the utility and robustness of such optical elements for broader quantum information processing applications can be substantial.
\end{abstract}


 \maketitle


\section{PHOTONIC QUANTUM ALGORITHMS WITHIN A VOLUME HOLOGRAM}
\label{sec:intro}  

It is the objective of this manuscript to argue that the construction of a small, lightweight, field deployable, inexpensive, low-dimensional, quantum computer is feasible using existing photonics technology.  It is well recognized that photons offer great promise in quantum information processing (QIP) given their robustness to decoherence.  However, it is this very resiliency that hinders their utility in quantum computing.  Their weak coupling with atomic structures leads to substantial inefficiencies. While they may not find themselves as the central facet  of a quantum CPU, they can, nevertheless, be an integrable part of the quantum information processing done by a quantum computer or in a quantum communication system where repetitive low-dimensional, fixed and robust  quantum algorithmic tasks are required (e.g. quantum error correction, quantum relays, quantum memory/CPU buss, ...).

Various approaches have been proposed for quantum computing.  Perhaps the most familiar of these is the quantum circuit model (QCM) \cite{NieChu00}.  Here the challenge is to identify a suitable subset of a universal set of quantum gates that can  faithfully represent a family of unitary operations on $d$ qubits ($d$-partite), such that the the number of gates required scale polynomially in $d$, whilst the dimension grows exponentially as $2^d$ .  After the unitary evolution the output needs to be projected onto the computational basis and irreversibly measured.  One realization of the QCM has been linear optical quantum computing (LOQC) and one-way quantum computing (OWQC), which are also known as cluster-state quantum computing \cite{KniLafMil01,RauBri01,Nie05,Kwi02}.  The essential feature of each of these approaches involves either a non-linear measuring process or a preparation of hyper-entangled input states ($|IN\rangle$), or both. Here a sequence of measurements are made to project the output state ($|OUT\rangle$)  onto one or another of the computational basis states, i.e onto a mutually unbiased basis (MUB).  In this manuscript we will explore the utility of phototnic quantum circuit models.

In this manuscript we suggest that in principle any quantum algorithm can be encoded within a single hologram, and that in practice, it has been found that one requires a $1 mm/N$ thickness of holographic material, where ($N$)  is the dimension of the state space.  This thickness will ensure high fidelity unitary transformations \cite{Gle10}.  Nevertheless, there are commercially available holographic media, e.g. OptiGrate's photo-thermal refractive (PTR) glass,  that can support faithful holograms with a thickness approaching a few centimeters, thereby extending  our considerations to 10 to 20 dimensional quantum state spaces \cite{CiaGleSmi06}.  Every application of LOQC to quantum gates that we are aware of requires a cascade of interferometers which is highly unstable and impractical.  The approach here will ``lock"  these interferometers within a tempered piece of glass that is resistant to environmental factors.

In this paper we will demonstrate that quantum gates and simple quantum algorithms can be efficiently simulated by the diffraction phenomena of a photon within a multiplexed holographic element.  In Sec.~\ref{sec:vhg} we briefly outline the structure and function of a volume holographic grating that will be referred to  throughout this manuscript.   In Sec.~\ref{sec:lm} we describe the photonic  quantum eigenstates we use.  In particular we use the space spanned by the photon's linear momentum (LM) as measured by the number of waves of tilt across the aperture, i.e. the transverse component of the wave vector parallel to the face of the hologram.
In Sec.~\ref{sec:sorter} we will briefly review our previous design of a quantum projection operator.
In Sec.~\ref{sec:qt} we discuss the two properties of quantum computing within the QCM that make this approach attractive. First, any conditional measurement can be commuted in time with any unitary quantum gate - the ``timeless nature of quantum computing." Second, photon entanglement can be encoded as a superposition state of a single photon in the higher- dimensional state spaces afforded by LM.  Our theoretical and numerical results indicate that OptiGrate's photo-thermal refractive (PTR) glass is an enabling technology.
Additionally, we give credence to the projection operator approach of Sec.~\ref{sec:sorter}
by applying it to a quantum teleportation (QT)  algorithm.  We choose this algorithm because it exhibits both quantum entanglement and the timeless nature of quantum computing \cite{Mil11,NieChu00}.  We used coupled-mode theory and supportive numerical simulations in our research, all with parameters consistent with PTR glass \cite{Mil11,Mil11b}. In Sec.~\ref{sec:cnot} we describe the construction of the familiar $CNOT$ gate without the need for multiplexing. Our alternative approach will ordinarily require stacking more holographic gratings. In conclusion, (Sec.~\ref{sec:sw}) we will  discuss the strengths (high efficiencies, robustness to environment) and limitations (scalability, crosstalk) of this technology. The holographic approach presented here is not scalable, and with existing technology it is not reprogrammable; nevertheless, the utility and robustness of such optical elements for broader quantum information processing applications can be substantial.  There should be many facets of a quantum computer other than the quantum CPU that could take advantage of this technology, including error correction, QKD relays, quantum memory management tasks and the like.

\section{VOLUME HOLOGRAPHIC GRATINGS} \label{sec:vhg}

Volume holography is used today for 2D image storage utilizing $394\, pixels/\mu m^2$, which consumes only $1\%$ of the theoretical volumetric storage density ($1/\lambda^3$) \cite{Bur01}.  This field was first introduced by Dennis Gabor in 1948 and it  has been a popular research field ever since the development of the laser in 1960.   Thicker holograms have more precise angular selectivity, i.e. its ability to differentiate the difference between two plane waves separated by a small angle,  and under certain well-known conditions  can achieve near perfect  efficiency \cite{Goo05}.  A  hologram is considered a volume hologram if its thickness $d \gg \Lambda^2/\lambda,$ where $\Lambda$ is the characteristic period of the index of refraction of the grating, and $\lambda$ is the wavelength of the light.  It is important for our purposes to emphasize that volume holography enables higher storage densities, and under suitable recording configurations can achieve near perfect efficiencies. This is not possible with thin holograms and  planar gratings (spatial light modulators) where efficiencies can be at most $33\,\%$.

\begin{figure}
   \begin{center}
   \begin{tabular}{c}
   \includegraphics[height=6cm]{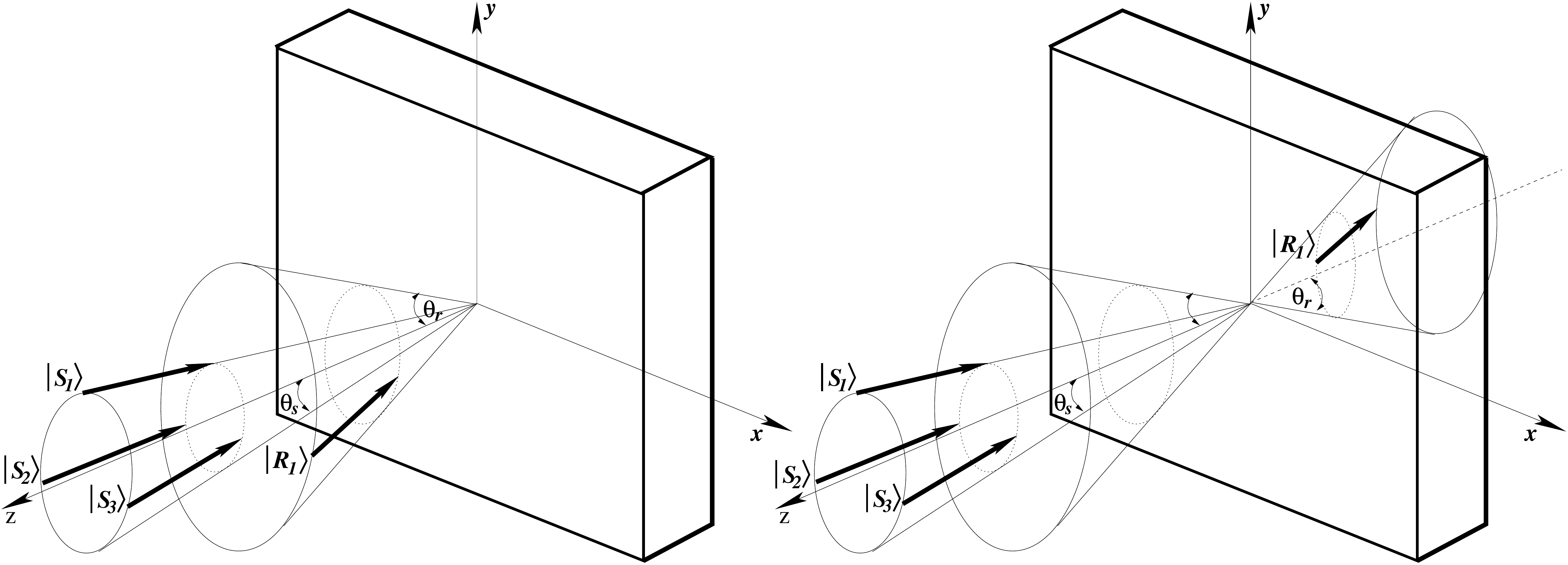}
   \end{tabular}
   \end{center}
   \caption[example]
   { \label{fig:vhg}
   The left diagram shows a  recording of a volume transmission grating by the coherent superposition of a plane reference wave
   $|R_1\rangle$ and a linear superposition of three signal waves
   $|S\rangle = e^{i \alpha_1} |S_1\rangle + e^{i \alpha_2} |\beta S_2\rangle + e^{i \alpha_3} | S_3 \rangle$ .
   On the right we show the function of the hologram. If the identically oriented signal wave $|S\rangle$ is sent
   into the hologram then the reference wave, $| R_1\rangle$, will be reconstructed in the diffraction.
   The diffraction pattern will ordinarily consist of higher order diffracted modes parallel to the signal state.
   However, for a suitably tuned volume hologram perfect efficiency can be achieved, as shown in the right diagram \cite{Mil11}.  This is why we  constrained the signal wave components to a cone of half angle $\theta_s$ centered on the normal to the hologram face. }
   \end{figure}

Briefly, the transmission volume holograms we consider here are  formed when a ``signal" wave, $\langle \vec r | S \rangle= A(\vec r) e^{i\Phi(\vec r)}$ is directed into a holographic material and made to coherently interfere with an oblique  "reference" plane wave, $|R\rangle$ as illustrated in the left diagram of  Fig.~\ref{fig:vhg} for  $N=3$.  In the figure the ``signal" wave is a superposition of $N$ plane waves,
\begin{equation}
\langle \vec r |S\rangle = \sum_{i=1}^N \langle \vec r | S_i \rangle =   \sum_{i=1}^N e^{i\alpha_i} e^{i \vec k_i \cdot \vec r},
\end{equation}
where $\alpha_i$ are pure phase angles.  In this paper,  we only consider planar reference waves,  and the signal state as the superposition of plane waves.  Ordinarily the signal waves will have variable phase and amplitude modulations.  After the hologram is developed, and if we direct the identical signal wave, $\langle \vec r | S\rangle$,  into the hologram, then for a perfectly tuned hologram,  the reference plane wave, $\langle \vec r | R\rangle$,  should emerge from  the hologram as illustrated in the right diagram of Fig.~\ref{fig:vhg}.  If it is not tuned, then other diffracted orders, e.g.  modes parallel to the signal states, may emerge.

How does one form a perfectly tuned hologram within the coupled mode theory? We have shown recently \cite{Mil11} that near perfect efficiencies can be obtained if (1) the hologram thickness is tuned to its optimal thickness, (2) if the each of the signal's Fourier wave vectors have the same projection onto the normal to the hologram surface, i.e. they all lie on a cone with half angle $\theta_s$ as shown in Fig.~\ref{fig:cone}, and (3) each of the reference plane waves lie on their own distinct cone concentric with the first, with half angle $\theta_r$ and centered on the normal to the hologram face.   We will also consider multiplex holograms wherein multiple independent exposures are made within the holographic material before it is developed.  We  demonstrated  using coupled-mode theory that if the  signal waves $\{S_i\}_{i,1,2, ...N}$  form an orthogonal set under the $L_2$ norm in the plane perpendicular to the waves propagation direction ($z$), i.e.
\begin{equation}
\langle S_i | S_j \rangle = \int S^*_i(x,y) S_j(x,y) dxdy = \delta_{ij},
\end{equation}
then  perfect efficiency can be achieved for each of the signals \cite{Mil11}.

A volume multiplexed hologram that has achieved perfect efficiency (within coupled-mode theory \cite{Kog69}) under the ``3+1"  conditions outlined above provides a linear map between signal and reference modes. Physically it represents a projection (or redirection) operator or signal state sorter,\cite{Mil11}
\begin{equation} \label{eq:po}
\hat{\cal  P} = | R_1 \rangle\langle S_1 | + | R_2 \rangle\langle S_2 | + \ldots + | R_N \rangle\langle S_N|,
\end{equation}
uniquely identifying each pair of signal and reference waves.

Although the index of refraction within the hologram can be rather complicated, these devices are strictly linear optical components.  Therefore, the diffraction patterns for a beam of photons will correspond exactly to the probability distribution for a single photon in the beam.  For the remainder of this manuscript we will assume we are dealing with low number Fock states.  We will describe in the next two sections (1) how we encode a $d$-partite quantum state onto a single photon that can be used in volume holography, and (2) how the unitary operation representing an entire quantum algorithm can be encoded in a multiplexed hologram.

\section{ENCODING A $D$-PARTITE STATE ON A SINGLE PHOTON} \label{sec:lm}

The dimension, $N$, of a quantum state space spanned by the direct product of $d$ qubits grows exponentially, $N=2^d$.  Thus to represent a single $d$-partite state by a single photon requires vastly more quantum numbers than available to its two polarization degrees of freedom.  However, the photon can be characterized by its extrinsic properties as characterized by its wavefront  amplitude and phase. The potential of extending photon-based quantum information processing and quantum computing to higher dimensions was made possible in 1983 when Miller and Wheeler \cite{MilWhe84} described a fundamental  quantum experiment utilizing a photon's orbital angular momentum (OAM), and when Allen et~al \cite{All92}  described Laguerre-Gaussian light beams that possessed a quantized orbital angular momentum (OAM) of $l\hbar$ per photon.   This opened up an arbitrarily high dimensional quantum space to a single photon \cite{MilWhe84,AllBarPad03}.  Following these discoveries Mair et~al. \cite{MaiVazWeiZel01,Oem04} unequivocally demonstrated the quantum nature of photon OAM by showing that pairs of OAM photons can be entangled using the non-linear optical process of parametric down conversion. Shortly thereafter, Molina-Terriza et~al. \cite{MolTorTor02} introduced a scheme to prepare photons in multidimensional vector states of OAM commencing higher-dimensional QIP, with applications to quantum key distribution. Recently a practical method has been demonstrated to produce such MUB states using computer-generated holography with a single spatial light modulator (SLM) \cite{Gru08}.

While photons with specific values of OAM have been emphasized recently in the literature,  one can equally well utilize any other extrinsic set of orthogonal basis functions for higher-dimensional QIP, e.g. energy, linear momentum, angular momentum.  While OAM states respect azimuthal symmetry, transverse or linear momentum (LM) states respect rectilinear symmetry.  Independent of the representation we use, the mutually unbiased (MUB) states will ordinarily be modulated in both amplitude and phase\cite{Gru08}.

While the advantage of higher-dimensional QIP lies in its ability to increase bandwidth while simultaneously tolerating a higher bit error
rate (BER) \cite{Cer02},  such transverse photon wave functions are more fragile to decoherence under propagation than the photon's spin wave function \cite{Pat05}, and the divergence of the states in propagation may require larger apertures.  However, since the spatial distances within a quantum computer are small this is not a concern here.  This is not the case for quantum key distribution.  Our preliminary results show that the efficiencies one can achieve with holographic elements exposed by LM plane waves and their superpositions (volume Bragg gratings) is substantially higher than that for photon wavefronts exhibiting OAM \cite{Mil11,Mil11b}.  Multiplexed volume Bragg gratings have been thoroughly addressed in the literature \cite{Kog69,Cas75,CiaGleSmi06,MohGayMag80}.  Since the efficiencies are potentially higher, and volume Bragg gratings are well understood, we therefore will restrict our attention in this paper to the orthonormal transverse photon states of LM.

\begin{figure}
   \begin{center}
   \begin{tabular}{c}
   \includegraphics[height=7cm]{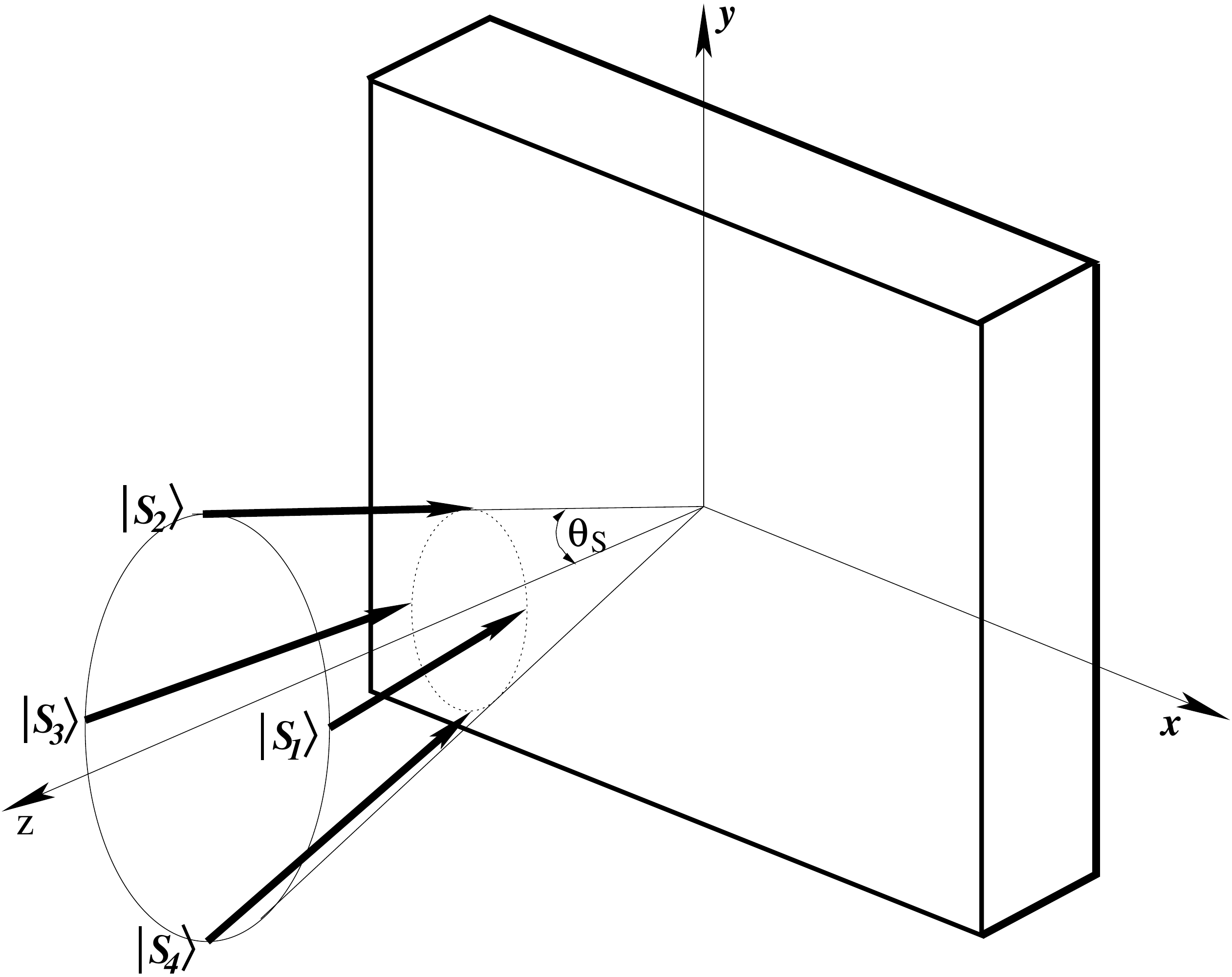}
   \end{tabular}
   \end{center}
   \caption[example]
   { \label{fig:cone}
  These Four LM states (arrows) that lie on a cone are used to generate a bipartite, or  4-dimensional signal state space.   Here we chose the eigenstates (each a plane wave) in this 8-dimensional space with eigenvectors, $|S_i\rangle = |\theta_s,\phi_i\rangle,\ i\in{1,2,3,4}$, here $\phi_1=0$, $\phi_2=\pi/2$, $\phi_3=\pi$ and $\phi_4=3\pi/2$.  This can be generalized to higher dimensions, $2^d$, by introducing $d$ eigenstates around the cone, as long as the angular selectivity of the hologram is less than $\Delta \phi=2\pi/2^d$. The wave vectors are defined in Eq.~\ref{eq:momentum}. }
   \end{figure}

We  concentrate in this section on the physics of transverse or LM states.  We quantize the LM within the  2-dimensional plane of the face of the hologram,  Fig.~\ref{fig:cone}.  We know that Bragg gratings can obtain perfect efficiencies as long as the component perpendicular to the face of the hologram  ($k_\perp$) for each of the photons' momentum are the same i.e. the plane wave states must lie on a cone.  In addition the thickness of the hologram must be carefully chosen \cite{Kog69,Goo05} .

If we are to represent a $d$-partite state by a single LM photon, we will require at least $N=2^d$ LM states around the cone of half angle $\theta_s$ centered on the normal to the hologram's face.  As a result, we can freely choose as our  computational quantum state basis any $2^d$ non-colinear plane waves. In this case, our integer quantum numbers will be the azimuthal angles, $\phi_i,\ i\in\{1,2,\ldots 2^d\}$ locating the wave vector of the photon around the cone.   Here we assume the hologram surface is in the $x$-$y$ plane.  For photons of wavenumber $k$ these waves correspond to a transverse linear momentum $p_\parallel =  \sin{(\theta_s)} \hbar k = \hbar k_\parallel$, with components $p^x_i\ =  \sin{(\phi_i)} \hbar k_\parallel$,    $p^y_i\ =  \cos{(\phi_i)} \hbar k_\parallel$ and $p^z_i  = \cos{(\theta)} \hbar k = \hbar k_\perp = constant$; respectively.  Here,  $k_\parallel = k \left( \lambda/D\right) $ is the magnitude of the transverse component of the wave vector of a plane wave of wavelength $\lambda$  with one "wave of tilt"  ($\lambda/D$) across the aperture of breadth $D$.

In the frame of the hologram,  and in units where the speed of light is
unity, the components of the 3-momentum, $\vec p = \{p^x, p^y, p^z\}$,  for each of our $2^d$ photons can be expressed in terms of their wavenumber, ($k$), i.e.,
\begin{equation}\label{eq:momentum}
\vec p_{i} =  \hbar \vec k_i =   \hbar \ \left\{k^x_i,k^y_i,k^z_i\right\} =
\hbar k \ \left\{\sin{(\theta)} \cos{(\phi_i)},\sin{(\theta)} \sin{(\phi_i)}, \cos{(\theta_s)}\right\}.
\end{equation}
These $N=2^d$ plane wave states, $\langle \vec r| S_i\rangle = e^{i \vec k_i \cdot \vec r}$   define our {\em computational basis} for  quantum information processing.
\begin{equation}
\label{eq:basis}
MUB = \left\{ | S_1 \rangle, | S_2 \rangle, \ldots,  | S_{N} \rangle \right\}.
\end{equation}
Each of these states represents a transverse Fourier mode of a photon;
they are orthogonal under the $L^2$ norm ($\langle S_i | S_j \rangle = \delta_{ij}$) and
span our $2^d$-dimensional state space.

\section{REDIRECTING STATES IN A COMPUTATIONAL BASIS FROM REFERENCE TO STATE SPACE}\label{sec:sorter}

In Sec.~\ref{sec:vhg} we outlined our recent progress in efficiently sorting a single photon  with
arbitrary complex wavefronts (Eq.~\ref{eq:po})\cite{Mil11,Mil11b}. And we showed that under certain conditions perfect efficiency) within coupled-mode theory) can be achieved.  We showed that a multiplexed hologram could function as a  $d$-partite
version of the more familiar qubit  polarizing beam splitter, i.e. a single optical element  that can efficiently sort each of the $d$-partite states in the $MUB$, or computational basis (Eq.~\ref{eq:basis}). As we argued, thick
holographic gratings fortunately produce high diffraction efficiency
in the first order \cite{Goo05,Kog69,Cas75}. If several
predominant diffracted orders are required, as is the case for
sorting, several independent fringe structures can coexist within the
holographic material.  Such multiplexed holograms have been used for multiple-beam
splitters and recombiners \cite{Cas75}, and more recently for
wide-angle beam steering\cite{Cia06}.  In a recent  paper we described
such a MUB-state sorter based on a multiplexed thick holographic
element constructed from commercially available photo thermal
refractive (PTR) glass \cite{Mil11}. Due to the unique properties
of PTR glass the grating's thickness can approach several {\it mm} and
be highly Bragg selective \cite{Efi99}. There is evidence that such sorters can be
highly efficient, with Bragg efficiency $> 95\%$ \cite{Cia06}.  Our recent simulations
 and empirical data on thick Bragg gratings indicate
that they may provide an adequate solution to this critical sorting and beam steering problem for ``single photon" QIP.

While the signal state sorter, Eq.~\ref{eq:po}, can be constructed in a multiplexed volume hologram under the strict conditions outlined in Sec.~\ref{sec:vhg}, we could just as well construct a multiplexed volume Bragg  transmission hologram, $\hat{\cal R}$, that recombines a beam. i.e. maps  the reference waves onto their corresponding signal states.   Mathematically that is the Hermitian conjugate of Eq.~\ref{eq:po},
\begin{equation} \label{eq:rdo}
\hat{\cal R} = \hat{\cal P}^\dag = | S_1 \rangle\langle R_1 | + | S_2 \rangle\langle R_2 | + \ldots + | S_n \rangle\langle R_n|.
\end{equation}
This can be constructed physically using the same hologram that we used for $\hat{\cal P}$, but instead of directing the signal waves to the hologram to reconstruct the corresponding reference wave, we could direct a reference wave which will reconstruct the desired superposition of signal wave states, i.e. the signal beam, $|S\rangle$.  For example,  two identical multiplexed holograms ($\hat{\cal P}$)  stacked back-to-back would represent the identity operator, $\hat{\cal I}$,  on this state space.  Any  signal wave used to record the hologram emerges unchanged  from the holographic sandwich; however, within the holographic material it will undergo a complicated evolution into the reference states, $|R_i\rangle$, and back.  Though, this identity gate may not be particularly useful in its own right, it does help illustrate our point.   We will find in the next two sections that these holographic state redirection elements (Eq.~\ref{eq:po}) can be useful in the construction of quantum computing gates.

\section{QUANTUM TELEPORTATION IN GLASS} \label{sec:qt}

To best illustrate how a quantum algorithm can be encoded within a single hologram we will focus our attention on quantum teleportation ($QT$).  This three qubit gate lives in an 8-dimensional state space.   It exhibits all the essential features we need to address, i.e. entanglement and conditional measurements.  The approach here will generalize, in principle,  to any quantum circuit model. We will adhere as closely as we can to the notation found in Nielsen and Chuang \cite{NieChu00}.

\begin{figure}
   \begin{center}
   \begin{tabular}{c}
   \includegraphics[height=4cm]{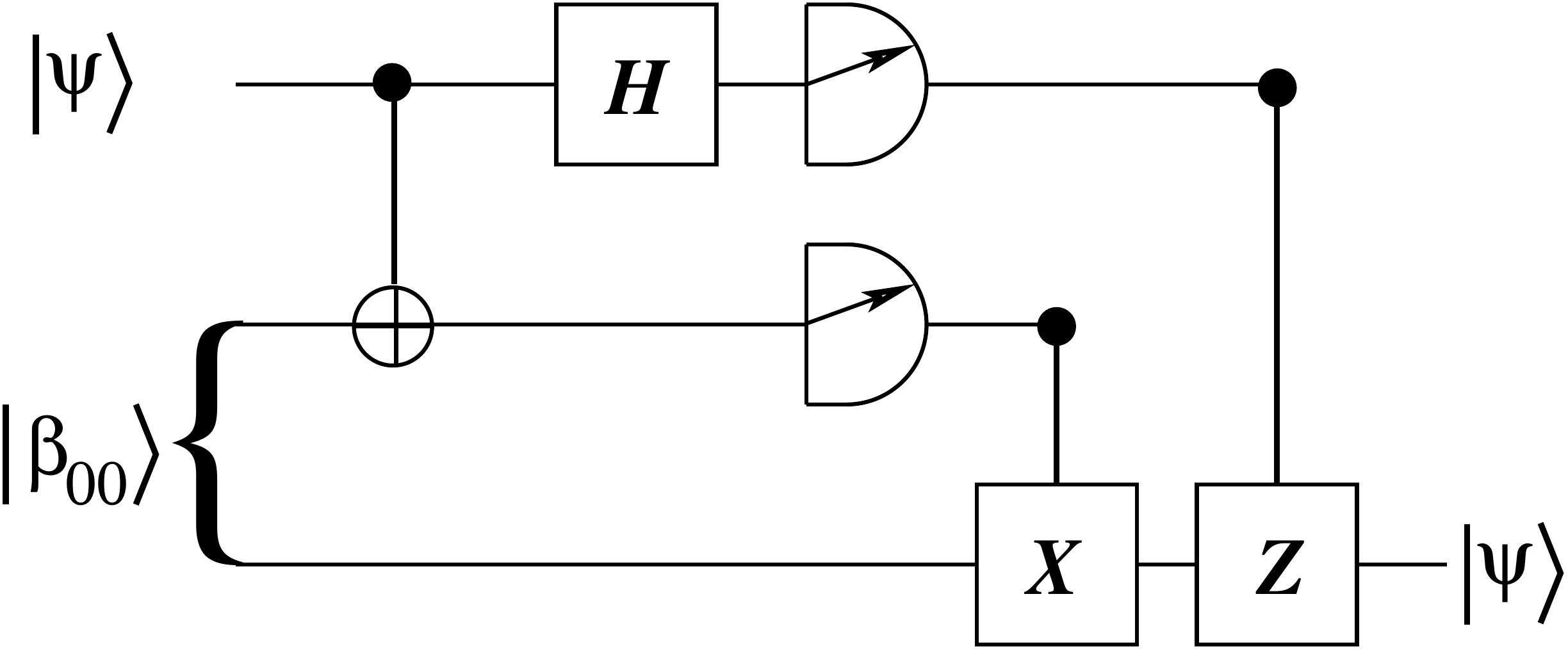}
   \end{tabular}
   \end{center}
   \caption[example]
   { \label{fig:qt}
   A quantum teleportation circuit \cite{NieChu00}. Here the 8-dimensional tripartite input state $|IN\rangle = |\psi\rangle \otimes |\beta_{00}\rangle$ is the direct product of the arbitrary qubit state, $|\psi\rangle=1/\sqrt{2} \left( |0\rangle + |1\rangle\right)$  we wish to teleport, and a maximally entangled pair of qubit states $|\beta_{00}\rangle=1/\sqrt{2} \left( |00\rangle + |11\rangle\right)$. This circuit is composed of an initial $CNOT$ operation, followed by a Hadamard gate, and two Pauli transformations ($X$ and $Z$) conditioned on two measurements.}
   \end{figure}

\begin{figure}
   \begin{center}
   \begin{tabular}{c}
   \includegraphics[height=5cm]{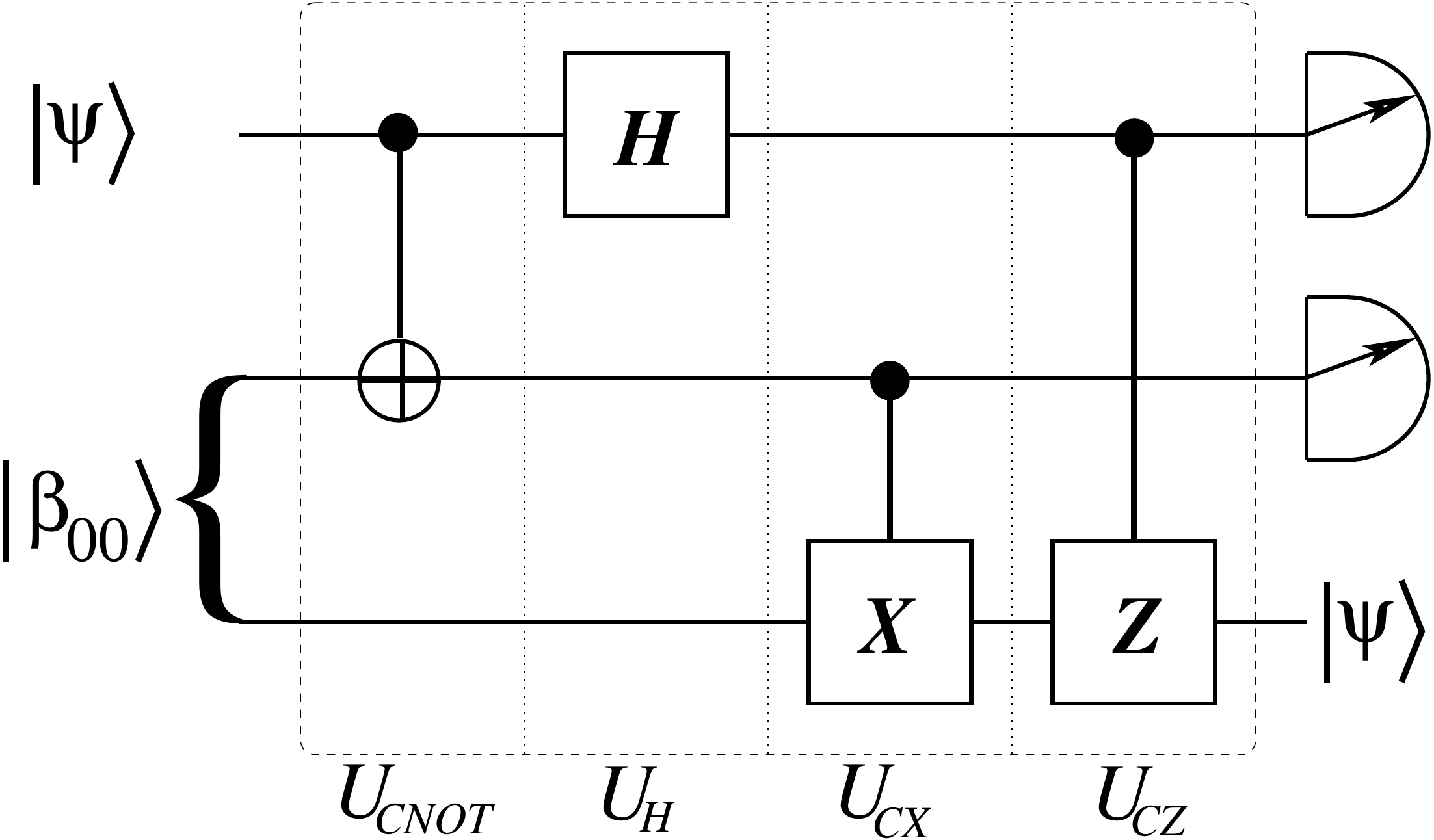}
   \end{tabular}
   \end{center}
   \caption[example]
   { \label{fig:qte}
   A equivalent quantum teleportation circuit \cite{NieChu00}. We use the ``principle of deferred measurement"  and commute the time ordering of the conditional measurements with the last two Pauli gates. This emphasizes the ``timeless'' feature of quantum information processing. The QT circuit thus becomes a single unitary transformation followed by two measurements and the transported state (top wire $\rightarrow$ bottom wire).}
   \end{figure}

The standard experimental realization of $QT$ utilizes a pair of entangled qubits, $|\beta_{00}\rangle$, and an arbitrary qu$b$it state vector to be teleported, $\psi$, where,
\begin{eqnarray}  \label{eq:bell}
|\beta_{00}\rangle &=& \frac{1}{\sqrt{2}} \left(  |00\rangle + |11\rangle    \right) \\
\psi &= &\alpha |0\rangle + \beta |1\rangle
\end{eqnarray}
The $QT$ circuit consists of a $CNOT$ gate, a Hadamard gate as well as two conditional measurements, one of which is  fed into a Pauli  $X$ gate, the other into a Pauli $Z$ gate as shown in Fig.~\ref{fig:qt}.  We exploit  the "timelessness" of quantum computing by using  the {\em principle of deferred measurements} and commute the two measurements in Fig.~\ref{fig:qt} with the quantum $X$ and $Z$ gates.  We arrive at an equivalent circuit model for $QT$ where all measurements are made in the end as illustrated in Fig.~\ref{fig:qte}.  We focus in this section on this equivalent tripartite $QT$ circuit.  Essentially $QT$ amounts to (1) preforming a unitary transformation on the input  tripartite state followed by, (2)  two measurements,
\begin{equation}
|OUT\rangle = \hat{\cal U}_{QT}\, |IN\rangle,
\end{equation}
where
\begin{equation}
\label{eq:input}
|IN\rangle = |\psi\rangle \otimes |\beta_{00}\rangle.
\end{equation}
This unitary transformation, $\hat{\cal U}_{QT}$, can be decomposed into a product of four sequential unitary transformations as illustrated by  the dashed box shown in Fig.~\ref{fig:qte}.  Evolution proceeds from the left to the right.  The input state first encounters (1)  a $CNOT$ gate on the first two qubits with unitary transformation, $\hat{\cal U}_{CNOT}$,  followed by (2) a Hadamard transformation on the first qubit with unitary operator, $\hat{\cal U}_H$, followed by (3) a Pauli $X$ gate on the third qubit conditioned on the second qubit with unitary transformation, $\hat{\cal U}_{CX}$, and finally (4) a Pauli $Z$ gate on the third qubit conditioned on the first qubit with unitary transformation, $\hat{\cal U}_{CZ}$.  We can therefore write the overall $QT$ unitary transformation as a product of these four transformations,
\begin{equation} \label{eq:uqt}
\hat{\cal U}_{QT} = \hat{\cal U}_{CZ}\,\,\hat{\cal U}_{CX}\,\,\hat{\cal U}_{H}\,\,\hat{\cal U}_{CNOT}
\end{equation}

This 8-dimensional state space can be interpreted as a direct product of 3 qubit states (Eq.~\ref{eq:input}) with entanglement in the last two qubits (Eq.~\ref{eq:bell}).  However, we can also represent the tripartite state as a pure state in an 8-dimensional state space without loss of generalization.  In this way the entire input state (Eq.~\ref{eq:input}) can be represented by a single photon in an 8-dimensional LM eigenspace as described in Sec.~\ref{sec:sorter}. We just need to define a correspondence between a state in the computational basis and one of the LM eigenstates (Eq.~\ref{eq:basis}). We make this identification below.

It is convenient for our analysis, and our application to holography, to represent the equivalent $QT$ algorithm in matrix notation.  Here we choose the standard representation of the qubits,
\begin{equation}
|0\rangle = \left(\begin{array}{c} 1 \\ 0 \end{array}\right) \ \ \hbox{and} \ \  |1\rangle = \left(\begin{array}{c} 0 \\ 1 \end{array}\right).
\end{equation}
The two input states can then be expressed as vectors in a  2 and 4 dimensional state space; respectively,
\begin{equation}
|\psi\rangle = \left( \begin{array}{c} \alpha \\ \beta \end{array}    \right),
\end{equation}
\begin{equation}
|\beta_{00}\rangle = \frac{1}{\sqrt{2}}\, \left( \begin{array}{c} 1 \\ 0 \\ 0 \\ 1 \end{array}    \right).
\end{equation}
Therefore the input state can be expressed as a vector in an 8-dimensional state space,
\begin{equation}\label{eq:in}
|IN\rangle = |\psi\rangle \otimes |\beta_{00}\rangle = \frac{1}{\sqrt{2}}\, \left( \begin{array}{c} \alpha \\ 0 \\ 0 \\ \alpha \\ \beta \\ 0 \\ 0 \\ \beta \end{array}    \right).
\end{equation}

Following the definitions of Sec.~\ref{sec:lm},  we can freely choose the following computational basis vectors in this 8-dimensional state space and their correspondence to each of eight ``signal" LM plane waves:
\begin{equation} \label{eq:cbasis}
\langle \vec r |S_1\rangle\! =\! e^{i\vec k_1\cdot \vec r} \Longleftrightarrow   \left( \begin{array}{c} 1 \\ 0 \\ 0 \\ 0 \\ 0 \\ 0 \\ 0 \\ 0 \end{array}    \right),\ \
\langle \vec r |S_2\rangle\! =\! e^{i\vec k_2 \cdot \vec r} \Longleftrightarrow   \left( \begin{array}{c} 0 \\ 1 \\ 0 \\ 0 \\ 0 \\ 0 \\ 0 \\ 0 \end{array} \right),\ \
\cdots , \ \
\langle \vec r |S_8\rangle\! =\! e^{i\vec k_8 \cdot \vec r} \Longleftrightarrow  \left( \begin{array}{c} 0 \\ 0 \\ 0 \\ 0 \\ 0 \\ 0 \\ 0 \\ 1 \end{array} \right).
\end{equation}
The quantum numbers of the ``signal" state,
\begin{equation}
|S_i\rangle = |k_i^\parallel\rangle
\end{equation}
represents the projection of the linear momentum of the photon onto the face of the hologram, i.e. the transverse components of the LM.  There would ordinarily be two components of this momentum, $|S_i\rangle = |k_i^\perp, k_i^\parallel\rangle$, hence two quantum numbers. However, here each of these eight ``signal" plane waves lie on a cone of half angle $\theta_s$.  Therefore, the first quantum number is constant for all states,
\begin{equation}
k^\perp_i=k^\perp=constant,\ \forall i.
\end{equation}
 We define the  transverse quantum numbers by equally spacing the states  around the cone's perimeter with $\phi_1 = 0$ and $\Delta \phi = \pi/4$.  This is a generalization to eight plane waves of the four wave configuration illustrated in Fig.~\ref{fig:cone}.  Therefore, one can say  each state is described by two quantum numbers, $\theta$ and $\phi$; however, we hold $\theta$ fixed so that our holograms can achieve perfect efficiency (within coupled mode theory) in modeling a given unitary transformation. This will be described below.

The input state (Eq.~\ref{eq:in}) is then a linear superposition of four of the eight computational basis vectors; and correspondingly, can be represented by a linear superposition of four plane wave state vectors,
\begin{equation}
|IN\rangle = \frac{1}{\sqrt{2}} \left( \alpha |S_1\rangle + \alpha |S_4\rangle + \beta |S_5\rangle + \beta |S_8\rangle\right).
\end{equation}
This state has a wavefront with both amplitude and phase variations.  In a similar vein we can define a corresponding set of eight ``reference"  plane waves that will be used to construct our holograms.  Here we define them as we did for the ``signal" waves, on a seperate concentric cone of half angle, $\theta_r \neq \theta_s $,
\begin{equation}\label{eq:rw}
\langle \vec r |R_{1}\rangle \! =\! e^{i\kappa_1\cdot r}, \ \
\langle \vec r |R_{2}\rangle \! =\! e^{i\kappa_2\cdot r}, \ \
\cdots, \ \
\langle \vec r |R_{8}\rangle \! =\! e^{i\kappa_8\cdot r}.
\end{equation}
Similarly, each of these reference waves live on a second cone of half angle $\theta_r\ne \theta_s$ and are equally spaced, but offset with respect to the signal waves, around its perimeter with $\phi_1 = \pi$ and $\Delta \phi = \pi/4$.  It is important to note that the exact placement of the eigenstates around the cones and relative relationships between the reference waves and the signal waves are unimportant within the couple-mode theory. Any configuration can yield perfect efficiency. However, in practice it is preferable to make the angular separation between the  signal and reference waves as large as possible, since this can lead to higher angular selectivity and higher efficiencies. This is not a concern in this paper.

We can use these basis vectors to construct each of the four unitary transformations shown in the dashed boxes of Fig.~\ref{fig:qte}, thereby yielding a representation of the overall $QT$ unitary transformation, $\hat{\cal U}_{QT}$, given by Eq.~\ref{eq:uqt}. In particular we find,
\begin{equation} \label{eq:ch}
\hat{\cal U}_{CNOT} = \left(
\begin{array}{cccccccc}
1 & 0 & 0 & 0 & 0 & 0 & 0 & 0 \\
0 & 1 & 0 & 0 & 0 & 0 & 0 & 0 \\
0 & 0 & 1 & 0 & 0 & 0 & 0 & 0 \\
0 & 0 & 0 & 1 & 0 & 0 & 0 & 0 \\
0 & 0 & 0 & 0 & 0 & 0 & 1 & 0 \\
0 & 0 & 0 & 0 & 0 & 0 & 0 & 1 \\
0 & 0 & 0 & 0 & 1 & 0 & 0 & 0 \\
0 & 0 & 0 & 0 & 0 & 1 & 0 & 0
\end{array}
\right),
\ \ \
\hat{\cal U}_{H} = \frac{1}{\sqrt{2}} \left(
\begin{array}{cccccccc}
1 & 0 & 0 & 0 & 1 & 0 & 0 & 0 \\
0 & 1 & 0 & 0 & 0 & 1 & 0 & 0 \\
0 & 0 & 1 & 0 & 0 & 0 & 1 & 0 \\
0 & 0 & 0 & 1 & 0 & 0 & 0 & 1 \\
1 & 0 & 0 & 0 & -1 & 0 & 1 & 0 \\
0 & 1 & 0 & 0 & 0 & -1 & 0 & 1 \\
0 & 0 & 1 & 0 & 1 & 0 & -1 & 0 \\
0 & 0 & 0 & 1 & 0 & 1 & 0 & -1
\end{array}
\right),
\end{equation}
as well as,
\begin{equation} \label{eq:xz}
\hat{\cal U}_{CX} = \left(
\begin{array}{cccccccc}
1 & 0 & 0 & 0 & 0 & 0 & 0 & 0 \\
0 & 1 & 0 & 0 & 0 & 0 & 0 & 0 \\
0 & 0 & 0 & 1 & 0 & 0 & 0 & 0 \\
0 & 0 & 1 & 0 & 0 & 0 & 0 & 0 \\
0 & 0 & 0 & 0 & 1 & 0 & 0 & 0 \\
0 & 0 & 0 & 0 & 0 & 1 & 0 & 0 \\
0 & 0 & 0 & 0 & 0 & 0 & 0 & 1 \\
0 & 0 & 0 & 0 & 0 & 0 & 1 & 0
\end{array}
\right),
\ \ \
\hat{\cal U}_{CZ} = \left(
\begin{array}{cccccccc}
1 & 0 & 0 & 0 & 0 & 0 & 0 & 0 \\
0 & 1 & 0 & 0 & 0 & 0 & 0 & 0 \\
0 & 0 & 1 & 0 & 0 & 0 & 0 & 0 \\
0 & 0 & 0 & 1 & 0 & 0 & 0 & 0 \\
0 & 0 & 0 & 0 & -1 & 0 & 0 & 0 \\
0 & 0 & 0 & 0 & 0 & -1 & 0 & 0 \\
0 & 0 & 0 & 0 & 0 & 0 & -1 & 0 \\
0 & 0 & 0 & 0 & 0 & 0 & 0 & -1
\end{array}
\right).
\end{equation}
Therefore, in our computational basis the unitary transformation representing the QT can be obtained by substituting Eqs.~\ref{eq:ch}-\ref{eq:xz} into Eq.~\ref{eq:uqt},
\begin{equation}
\label{eq:muqt}
\hat{\cal U}_{QT} = \frac{1}{\sqrt{2}} \left(
\begin{array}{cccccccc}
1 & 0 & 0 & 0 & 0 & 0 & 1 & 0 \\
0 & 1 & 0 & 0 & 0 & 0 & 0 & 1 \\
0 & 0 & 0 & 1 & 0 & 1 & 0 & 0 \\
0 & 0 & 1 & 0 & 1 & 0 & 0 & 0 \\
-1 & 0 & 0 & 0 & 0 & 0 & 1 & 0 \\
0 & -1 & 0 & 0 & 0 & 0 & 0 & 1 \\
0 & 0 & 0 & -1 & 0 & 1 & 0 & 0 \\
0 & 0 & -1 & 0 & 1 & 0 & 0 & 0
\end{array}
\right).
\end{equation}
This unitary transformation of the equivalent $QT$ circuit shown in Fig.~\ref{fig:qte} accomplishes the desired transformation,
\begin{equation}
|IN\rangle = \!\!\underbrace{\psi}_{\begin{array}{c}   top \\  wire\end{array}} \! \otimes \!\!
\underbrace{|\beta_{00}\rangle}_{\begin{array}{c} bottom \\ two\, wires\end{array}} \!\!
\Longrightarrow \ \ \
 |OUT\rangle = \hat{\cal U}_{QT}\, |IN\rangle = \underbrace{ \left( \frac{|1\rangle-|0\rangle}{\sqrt{2}} \right)}_{\begin{array}{c} top\, wire\end{array}} \otimes
 \underbrace{\left( \frac{|1\rangle+|0\rangle}{\sqrt{2}} \right)}_{\begin{array}{c} middle\, wire\end{array}} \otimes \!\!
  \underbrace{\psi  }_{\begin{array}{c} bottom \\ wire \end{array}},
\end{equation}
thereby teleporting the state function from the top wire to the bottom wire.

We describe how to encode the unitary transformation, $\hat{\cal U}_{QT}$  into a single holographic element. By construction, the rows of $\hat{\cal U}_{QT}$ form an orthonormal basis that spans the 8-dimensional state space. Each of these rows can be expressed as a superposition of the computational basis state vectors of Eq.~\ref{eq:cbasis}.  For this example, each row requires a superposition of exactly two of the eight computational basis state vectors.   Therefore, each row can be represented as a linear superposition of the corresponding (Eq.~\ref{eq:cbasis}) plane wave states -- two plane waves each for this particular quantum circuit. For example, if we consider the state vector representing each of the  rows of the matrix in Eq.~\ref{eq:muqt}, we can construct each of the  the eight  $QT$ signal states, $|\Sigma_i\rangle$ that we will use in the successive recordings in the hologram,
\begin{eqnarray}
\label{eq:s1}
|\Sigma_1\rangle = \frac{1}{\sqrt{2}} \left(|S_7\rangle + |S_1\rangle\right)\ \  \Longleftrightarrow\ \  {U^\dag_{QT}}_{1,j} \\
|\Sigma_2\rangle = \frac{1}{\sqrt{2}} \left(|S_8\rangle + |S_2\rangle\right) \ \  \Longleftrightarrow\ \   {U^\dag_{QT}}_{2,j} \\
|\Sigma_3\rangle = \frac{1}{\sqrt{2}} \left(|S_6\rangle + |S_4\rangle\right)\ \   \Longleftrightarrow\ \   {U^\dag_{QT}}_{3,j} \\
|\Sigma_4\rangle = \frac{1}{\sqrt{2}} \left(|S_5\rangle + |S_3\rangle\right)\ \   \Longleftrightarrow\ \   {U^\dag_{QT}}_{4,j} \\
|\Sigma_5\rangle = \frac{1}{\sqrt{2}} \left(|S_7\rangle - |S_1\rangle\right) \ \  \Longleftrightarrow\ \   {U^\dag_{QT}}_{5,j} \\
|\Sigma_6\rangle = \frac{1}{\sqrt{2}} \left(|S_8\rangle - |S_2\rangle\right) \ \  \Longleftrightarrow\ \   {U^\dag_{QT}}_{6,j} \\
|\Sigma_7\rangle = \frac{1}{\sqrt{2}} \left(|S_6\rangle - |S_4\rangle\right)\ \   \Longleftrightarrow\ \   {U^\dag_{QT}}_{7,j} \\
\label{eq:s8}
|\Sigma_8\rangle = \frac{1}{\sqrt{2}} \left(|S_5\rangle - |S_3\rangle\right)\ \   \Longleftrightarrow\ \   {U^\dag_{QT}}_{8,j}.
\end{eqnarray}
By, inspection we can reconfirm that these eight basis vectors form an orthonormal basis spanning the 8-dimensional state space.

\begin{figure}
   \begin{center}
   \begin{tabular}{c}
   \includegraphics[height=4in]{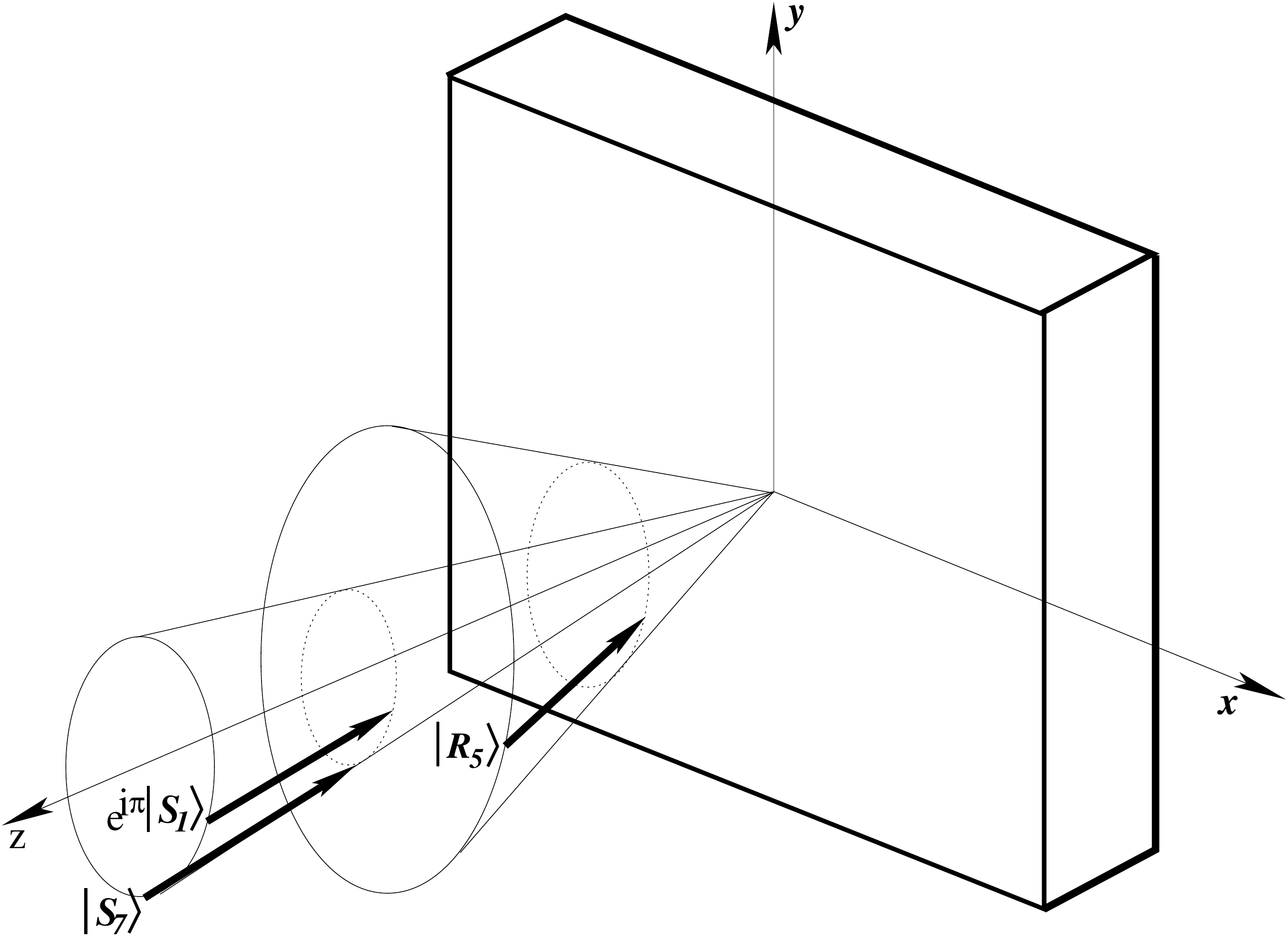}
   \end{tabular}
   \end{center}
   \caption[example]
   { \label{fig:qt5}
Volume holographic design of the fifth of eight independent  recordings used to generate the QT unitary transformation,  $\hat{\cal U}_{QT}$, in PTR glass.}
   \end{figure}

We construct a coordinate system where the $z$-axis is orthogonal to the face (the $x$-$y$ plane) of the hologram.  A faithful holographic representation of   $\hat{\cal U}_{QT}$ can be achieved by multiplexing eight recordings  within the holographic material.  The hologram is recorded so that each row of the unitary matrix, $\hat{\cal U}_{QT}$,  is used to generate its  own volume holographic grating.  A qubit gate such as the Hadamard gate ($H$) is represented by a $2\times2$ matrix and would ordinarily require two recordings.  A  2-qubit gate  such as the controlled-not gate $CNOT$ is represented by a $4\times 4$ matrix and would ordinarily require four independent recordings (this will be discussed in the next section).  The qutrit $QT$ circuit is represented by a $8\times8$ dimensional matrix (Eq.~\ref{eq:muqt}) and requires the following eight sets of independent recordings, between each of the ``signal" states in Eqs.~\ref{eq:s1}-\ref{eq:s8} and its corresponding ``reference"  plane wave of Eq.~\ref{eq:rw} as follows:
\begin{enumerate}
\item The first recording is made by the coherent superposition of ``signal state $\langle \vec r | \Sigma_1\rangle$ and  $\langle \vec r |R_1\rangle$;
\item The second recording is made by the coherent superposition of ``signal state $\langle \vec r | \Sigma_2\rangle$ and  $\langle \vec r |R_2\rangle$;
\item [] \centerline{\vdots}
\vskip -.75 cm
\item The eighth and final recording is made by the coherent superposition of ``signal state $\langle \vec r | \Sigma_8\rangle$ and  $\langle \vec r |R_8\rangle$
\end{enumerate}
The recording geometry for the case of the fifth recording is illustrated in Fig.~\ref{fig:qt5}.

Since (1) each of the eight signal states lie on a single cone with axis normal to the face of the hologram and form an orthonormal basis spanning the state space, and (2)  each of the eight reference states lie on another concentric cone and span an isomorphic 8-dimensional state space, the oct-tuple multiplexed hologram (within coupled mode theory) faithfully represents the desired transformation,
\begin{eqnarray*}\label{eq:qtsr}
\hat{\cal U}_{QT} = & \frac{1}{\sqrt{2}} \left(
|R_1\rangle\langle S_1| + |R_1\rangle\langle S_7| +
|R_2\rangle\langle S_2| + |R_2\rangle\langle S_8| +
|R_3\rangle\langle S_4| + |R_3\rangle\langle S_6| +
|R_4\rangle\langle S_3| + |R_4\rangle\langle S_5| -\right.\\
&\left. |R_5\rangle\langle S_1| + |R_5\rangle\langle S_7| -
|R_6\rangle\langle S_2| + |R_6\rangle\langle S_8| -
|R_7\rangle\langle S_4| + |R_7\rangle\langle S_6| -
|R_8\rangle\langle S_3| + |R_8\rangle\langle S_5|
\right).
\end{eqnarray*}
However, this operator maps quantum states in `signal'  space (inner cone of Fig.~\ref{fig:qt5}) onto corresponding states in the reference  space (outer cone of Fig.~\ref{fig:qt5}). In order to map this output back into the corresponding  states in the signal we would need to take the diffracted output of the hologram described by Eq.~\ref{eq:qtsr} as input to a ``redirection"  as described by Eq.~\ref{eq:rdo} in Sec.~\ref{sec:lm}.  This can be constructed in a similar fashion by eight recordings, wherein the $\hbox{i}^{th}$ recording is made by the coherent superposition of the $\hbox{i}^{th}$ signal plane wave basis state  $\langle \vec r | S_i \rangle = e^{i \vec k_i \cdot \vec r}$  with the $\hbox{i}^{th}$ reference plane wave basis state, $\langle \vec r | R_i \rangle = e^{i \vec \kappa_i \cdot \vec r}$, yielding,
\begin{equation}
\hat{\cal R}_{QT} = \sum_{i=1}^{8} |S_i\rangle\langle R_i|.
\end{equation}
Stacking the redirection hologram behind the hologram for the desired transformation will yield the desired QT operation.
\begin{eqnarray*}
\widehat{QT}  =  \hat{\cal R}_{QT}\,\, \hat{\cal U}_{QT} = & \frac{1}{\sqrt{2}} \left(
|S_1\rangle\langle S_1| + |S_1\rangle\langle S_7| +
|S_2\rangle\langle S_2| + |S_2\rangle\langle S_8| +
|S_3\rangle\langle S_4| + S_3\rangle\langle S_6| +
|S_4\rangle\langle S_3| + |S_4\rangle\langle S_5| -\right.\\
&\left. |S_5\rangle\langle S_1| + |S_5\rangle\langle S_7| -
|S_6\rangle\langle S_2| + |S_6\rangle\langle S_8| -
|S_7\rangle\langle S_4| + |S_7\rangle\langle S_6| -
|S_8\rangle\langle S_3| + |S_8\rangle\langle S_5|
\right)
\end{eqnarray*}

\section{THE CNOT GATE: STACKING INSTEAD OF MULTIPLEXING} \label{sec:cnot}

\begin{figure}
   \begin{center}
   \begin{tabular}{c}
   \includegraphics[height=4in]{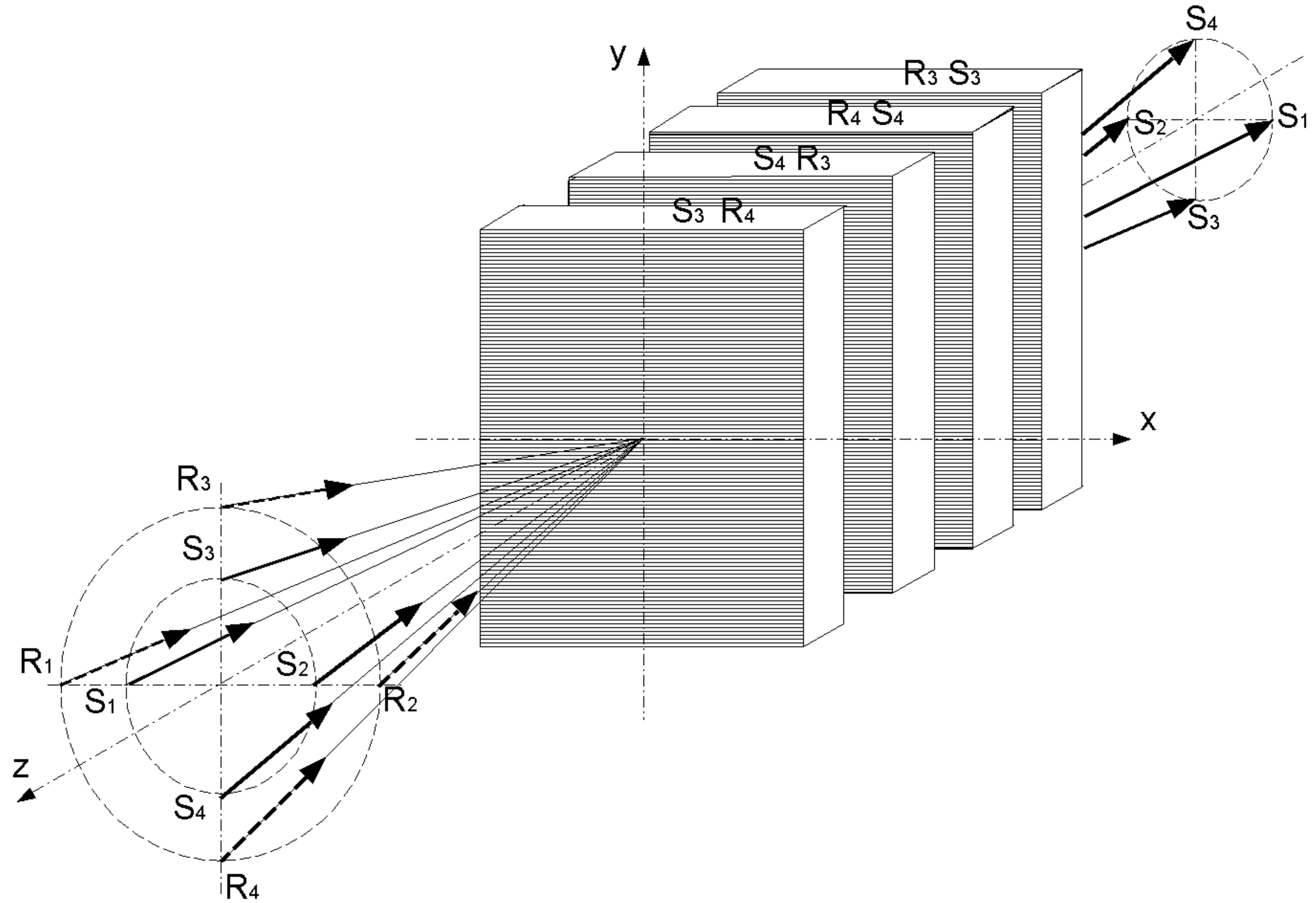}
   \end{tabular}
   \end{center}
   \caption[example]
   { \label{fig:cnot}
Volume holographic design of the 4-dimensional CNOT gate in PTR glass.  The gate can be constructed by a stack of 4 LM gratings, or by a stack of two multiplexed gratings. }
   \end{figure}

One alternative to multiplexing is to make single recordings in each of many holograms and to stack the holograms.  Here we provide a second illustrative example with a specific design of a quantum $CNOT$ gate compatible with  PTR glass.  This gate is realized by stacking four holograms, and we will describe each below.  The $CNOT$ gate is a two qubit gate. Therefore the dimension of the state space is  4-dimensional. While this state space can be constructed as a product space of qubits by utilizing the polarization states of two correlated photons, it can also be represented by a single LM photon in a 4-dimensional state space.  The  $CNOT$ gate can be constructed with a single photon.  Following the arguments of the previous section, we freely choose four independent plane waves lying on the cone shown in Fig.~\ref{fig:cnot}.  We associate to these independent transverse LM modes the four orthogonal quantum state vectors, $|S_1\rangle$, $|S_2\rangle$, $|S_3\rangle$ and $|S_4\rangle$.  Any state vector, $|\psi\rangle$, in this  4-dimensional state space can be written as a linear superposition of these states,
\begin{equation}
|\psi\rangle = \alpha |S_1\rangle + \beta |S_2\rangle + \gamma |S_3\rangle + \delta |S_4\rangle,
\end{equation}
with,
\begin{equation}
|\alpha|^2 + |\beta|^2+|\gamma|^2 + |\delta|^2 = 1.
\end{equation}
Each of our basis states can be expressed in matrix notation,
\begin{equation}
|S_1\rangle = \left( \begin{array}{c} 1 \\ 0 \\ 0\\ 0 \end{array} \right), \
|S_2\rangle = \left( \begin{array}{c} 0 \\ 1 \\ 0\\ 0 \end{array} \right), \
|S_3\rangle = \left( \begin{array}{c} 0 \\ 0 \\ 1\\ 0 \end{array} \right), \  \hbox{and}\
|S_4\rangle = \left( \begin{array}{c} 0 \\ 0 \\ 0\\ 1 \end{array} \right).
\end{equation}
In this computational basis the CNOT gate can be expressed by the following unitary transformation:
\begin{equation}
CNOT = \left( \begin{array}{c c c c}
1 & 0 & 0 & 0 \\
0 & 1 & 0 & 0 \\
0 & 0 & 0 & 1 \\
0 & 0 & 1 & 0
 \end{array} \right).
\end{equation}
If we let the $z$-axis be orthogonal to the face ($x$-$y$ plane) of the hologram, the four volume holographic gratings are recorded by a suitable superposition of the set of four signal plane waves,
\begin{equation}
\langle  \vec r | S_1\rangle = \exp\left(i \vec k_1 \cdot \vec r \right), \ \langle  \vec r | b\rangle = \exp\left(i k_2 \cdot \vec r \right), \
\langle  \vec r | S_2\rangle = \exp\left(i \vec k_3 \cdot \vec r \right), \  and \ \langle  \vec r | d\rangle = \exp\left(i k_4 \cdot \vec r \right),
\end{equation}
and four reference waves,
\begin{equation}
\langle  \vec r | R_1\rangle = \exp\left(i \vec \kappa_1 \cdot \vec r \right), \ \langle  \vec r | R_2\rangle = \exp\left(i \vec \kappa_2 \cdot \vec r \right), \
\langle  \vec r | R_3\rangle = \exp\left(i \vec \kappa_3 \cdot \vec r \right), \  and \ \langle  \vec r | R_4\rangle = \exp\left(i \vec \kappa_4 \cdot r \right),
\end{equation}
as shown in Fig.~\ref{fig:cnot}.

The hologram is recorded so that each row of the unitary matrix of the $CNOT$ gate  is used to generate its  own volume holographic grating.  For a 2-qubit gate such as the $CNOT$ gate we would ordinarily require four recordings; however, since the first two bits are just an identity matrix we need only two layers to transform the signal states into the desired reference states. In addition to one holographic recording per dimension of the state space,  we also require the conjugate of the each grating (two in the case of the CNOT gate) in order  to transform the diffracted reference waves from the reference waves back into the desired signal states. In particular, the CNOT-gate constructed from four holographic gratings stacked together as is shown in Fig.~\ref{fig:cnot}.
\begin{enumerate}
\item The first grating is recorded with the two coherent plane waves corresponding to states $|S_3\rangle$ and  $|R_4\rangle$ .
\item The second grating is recorded with the two coherent plane waves corresponding to states $|S_4\rangle$ and  $|R_3\rangle$ .
\item The third grating is recorded with the two coherent plane waves corresponding to states $|R_4\rangle$ and  $|S_4\rangle$ .
\item The fourth grating is recorded with the two coherent plane waves corresponding to states $|R_3\rangle$ and  $|S_3\rangle$ .
\end{enumerate}
The four gates will not diffract the first two signal states $|S_1\rangle$ or $|S_2\rangle$. However, the first two gratings redirect the two signal states $|S_3\rangle$ and $|S_4\rangle$ into $|R_4\rangle$ and $|R_3\rangle$ respectively, in  accordance with the Pauli $X$-gate,
\begin{equation}
\hat X = \left(  \begin{array}{c c} 0 & 1 \\ 1 & 0 \end{array}\right).
\end{equation}
The first hologram is equivalent to the operator,
\begin{equation}
\hat{\cal U}_{1} = |S_1 \rangle\langle S_1| + |S_2\rangle\langle S_2|+|R_4\rangle\langle S_3| + |S_4\rangle\langle S_4|,
\end{equation}
and the second hologram recorded with signal plane wave, $\langle \vec r | S_3 \rangle$ and reference plane wave,
$\langle \vec r | R_4 \rangle$, is equivalent to the operator,
\begin{equation}
\label{eq:cnotu2}
\hat{\cal U}_{2} = |S_1 \rangle\langle S_1| + |S_2\rangle\langle S_2|+|R_4\rangle\langle R_4| + |R_3\rangle\langle S_4|.
\end{equation}
After the signal state , $|IN\rangle$, passes through the first hologram the basis vectors change from $\{S_1, S_2, S_3, S_4\}$ to the orthonormal  basis, $\{S_1, S_2, R_4, S_4\}$. This explains why the third term in Eq.~\ref{eq:cnotu2} is entirely within the reference space.  

While these two recordings could have been made in a single multiplexed hologram, we recover the same function by stacking the two together, thereby generating the $CNOT$ operation,
\begin{equation}
\hat{\cal U}_{CNOT} = \hat{\cal U}_{2} \hat{\cal U}_{1} =  |S_1 \rangle\langle S_1| + |S_2\rangle\langle S_2|+|R_3\rangle\langle S_4| + |R_4\rangle\langle S_3|.
\end{equation}
However, the output of these two stacked holograms are the reference states, $|R_1\rangle$,  $|R_2\rangle$,  $|R_3\rangle$ and  $|R_4\rangle$. In order to redirect these back to the proper signal states, we require the redirection operator similar to Eq.~\ref{eq:rdo}. This can be accomplished by recording a third hologram with the states $|R_3\rangle$ and $|S_3\rangle$. The third hologram is equivalent to the operator,
\begin{equation}
\hat{\cal U}_{3} = |S_1 \rangle\langle S_1| + |S_2\rangle\langle S_2|+|S_3\rangle\langle R_3| + |R_4\rangle\langle R_4|.
\end{equation}
Similarily, the fourth hologram is recorded with the states $|R_4\rangle$ and $|S_4\rangle$ and the fourth hologram is equivalent to operator,
\begin{equation}
\hat{\cal U}_{4} = |S_1 \rangle\langle S_1| + |S_2\rangle\langle S_2|+|S_3\rangle\langle S_3| + |S_4\rangle\langle R_4|.
\end{equation}
Therefore, the combination of the four stacked volume holograms have the desired action -- the CNOT gate.
\begin{equation}
\widehat{CNOT} = \hat{\cal U}_4 \cdot \hat{\cal U}_3\cdot\hat{\cal U}_2 \cdot \hat{\cal U}_1 = |S_1 \rangle\langle S_1| + |S_2\rangle\langle S_2|+|S_4 \rangle\langle S_3| + |S_3\rangle\langle S_4|.
\end{equation}
One can apply these principles to design a universal set of quantum gates, as well as simple quantum algorithms such as QT.

The advantage of stacking the holograms is that one can make the hologram thicker, thereby increasing the efficiency; however, achieving and maintaining the proper alignment should be more problematic. By multiplexing, we would need two holograms, each with two independent recordings in them.  The first would be equivalent to the last two holograms in Fig.~\ref{fig:cnot}, while the second would be equivalent to the first two and would just redirect the reference beams into their corresponding signal states. The first two recordings are complementary to the second two -- thus in some sense we are recording the ``square root" of the CNOT gate.

\section{DISCUSSION: ADVANTAGES AND LIMITATIONS} \label{sec:sw}

Constructing simple quantum algorithms and quantum gates in volume holograms provides substantially greater optical stability  than the equivalent optical bench realization.  For LOQC, this stability is the overarching advantage.  Often quantum operators, e.g. the simple projection operator given by  Eq.~\ref{eq:po}, require a cascade of interferometers where the output of one interferometer is used as  the input of the next \cite{Kwi02,Lea04, Zou05}. Therefore, as the dimension of each state space increases,  it becomes exceedingly hard to stabilize and is simply impractical beyond two qubits.  Other approaches, such as crossed thin gratings lack the efficiency needed for QIP. The device proposed here can potentially achieve this in a single piece of glass without the problem of misalignment.  The technology presented here can potentially replace "fixed" optical components for a broad spectrum of classical and quantum photonics experiments.

The primary limitation of volume holographic QIP is that it is not scalable.  Experience shows that multiplexing requires approximately $1 mm$ per recording of the state space to achieve high fidelity, and in QIP applications this scales exponentially with the number of qubits. But then again, we are not aware of a realistic scalable quantum computer to date. Secondly,  the holograms discussed here are write-once holograms and cannot be erased. Therefore the algorithm is ``fixed" into the holographic element.  While there are re-recordable holographic media, none that we know of has the specifications to outperform PTR glass for the applications discussed in this manuscript.   This is hardly a prescription for a quantum CPU; however, as mentioned in the paper, this technology might be integral to complete QIP systems where smaller $d$-partite operations are needed on a routine basis, e.g. a quantum memory bus, quantum error correction circuit, QKD relay system, etc...

While we have extensively analyzed volume holographic gates and algorithms  using coupled-mode theory, paraxial wave equation simulations and finite-difference time domain simulations, we have not analyzed the engineering particulars of this device. For example: (1) how many independent writes of orthogonal states into a holographic element can be made in the PTR glass before cross-talk or saturation between the modes becomes a limiting factor? (2) Is it difficult to stack the holograms due to the enhanced angular selectivity of the volume holograms? And (3) what is the maximum number of recordings in a multiplexed PTR hologram that can be reasonably be achieved? In this sense, we are well along in understanding these devices from a theoretical  prospective; however, we are at the very beginning experimentally.

\acknowledgments     

WAM acknowledges support from AFRL/RITC under grant FA 8750-10-2-0017, and WAM and CT from the ARFL's Visiting Faculty and Student  Research Programs, as well as from the Complex Networks Program at the Air Force Office of Scientific Research. JM  acknowledges support from a National Research Council research associateship award at the AFRL/RITC, Rome, NY. Any opinions, findings and conclusions or recommendations expressed in this material are those of the author(s) and do not necessarily reflect the views of AFRL.




\end{document}